\documentclass[graybox]{svmult}

\usepackage{mathptmx}       
\usepackage{helvet}         
\usepackage{courier}        
\usepackage{type1cm}
\usepackage{subfigure}       
\usepackage{amsmath}
\usepackage{amssymb}
\usepackage{makeidx}         
\usepackage{graphicx}        
\usepackage{multicol}        
\usepackage[bottom]{footmisc}

\begin{document}

\title*{Statistical Properties of Fluctuations: A Method to Check Market Behavior}
\author{Prasanta K. Panigrahi, Sayantan Ghosh, P. Manimaran, Dilip P. Ahalpara}
\institute{Prasanta K. Panigrahi, Indian Institute of Science
Education and Research (Kolkata), Salt Lake City, Kolkata 700 106,
India and Physical Research Laboratory, Navrangpura,
Ahmedabad 380 009, India. \footnote{This paper is dedicated to the memory of Prof. J. C. Parikh, who was one of the founding fathers of econophysics in India.}\footnote{\texttt{prasanta@prl.res.in}} \\
\and Sayantan Ghosh, The Insitute of Mathematical Sciences, C.I.T.
Campus, Taramani, Chennai 600 113, India.\footnote{\texttt{sayantan@imsc.res.in}}\\
\and P. Manimaran, Center for
Mathematical Sciences, C R Rao Advanced Institute of Mathematics,
Statistics and Computer Science,
HCU campus, Hyderabad 500 046, India.\footnote{\texttt{rpmanimaran@gmail.com}}\\
\and Dilip P. Ahalpara, The
Insitute for Plasma Research, Bhat, Gandhinagar, 382 428, India.\footnote{\texttt{dilip@ipr.res.in}}}
%
%
\maketitle

\abstract{We analyze the Bombay stock exchange (BSE) price index
over the period of last 12 years. Keeping in mind the large
fluctuations in last few years, we carefully find out the transient,
non-statistical and locally structured variations. For that purpose,
we make use of Daubechies wavelet and characterize the fractal
behavior of the returns using a recently developed wavelet based
fluctuation analysis method. the returns show a fat-tail distribution as
also weak non-statistical behavior. We have also carried out
continuous wavelet as well as Fourier power spectral analysis to
characterize the periodic nature and correlation properties of the
time series.}

\section{Introduction}
Financial markets are known to show different behavior at different
time scales and under different socio-economic conditions. The
random behavior of fluctuations in the smaller time scales and the
manifestation of structured behavior at intermediate and long time
scales have been well studied \cite{plerou}-\cite{mandel}. Many of
the stock markets have shown large scale fluctuations during the
past three years. Here we concentrate on the behavior of the
fluctuation of the Bombay stock exchange (BSE) high price values in
daily trading. The point that makes the analysis of the BSE price index
interesting is the fact that it has a significant fluctuations on a
shorter time scale while growing tremendously over a longer time
period. The statistical properties of the fluctuations and the
behavior of the returns of such a growing market are of particular
interest. Wavelet transform \cite{daub}-\cite{ram} based
multi-resolution analysis \cite{mani1,mani3} has been successfully
used earlier to analyze time series from various areas
\cite{drozdz}-\cite{mani4}.

\noindent In this work, we analyze the BSE high price index value
using both continuous and discrete wavelet transform and
multifractal detrended fluctuation analysis (MF-DFA)
\cite{khu}-\cite{mohaved}. We use the continuous
wavelet transform (CWT) to analyze the behavior of the time series
at different frequencies and extract the periodic nature of the series
if existent. The discrete wavelet transform based method is used to
find the multifractal nature of the time series. For the purpose of
comparison, the MF-DFA method is used for characterization of the time series. It has been observed in \cite{mani2}
that BSE returns showed a Gaussian random behavior and certain non-statistical features.

The present work is organized as follows. Section \ref{sec:wt} contains a
brief description and applications of the continuous and discreet wavelet transforms.
The discrete wavelet based method \cite{mani4} to analyze fluctuations
is reviewed in section \ref{sec:wbm}. In section \ref{sec:obs}, the data is analyzed
through the wavelet based method, MF-DFA and Fourier analysis. We conclude in section
\ref{sec:disc} with results and a brief discussion.

\begin{figure}
\includegraphics[width=11cm,height=6cm]{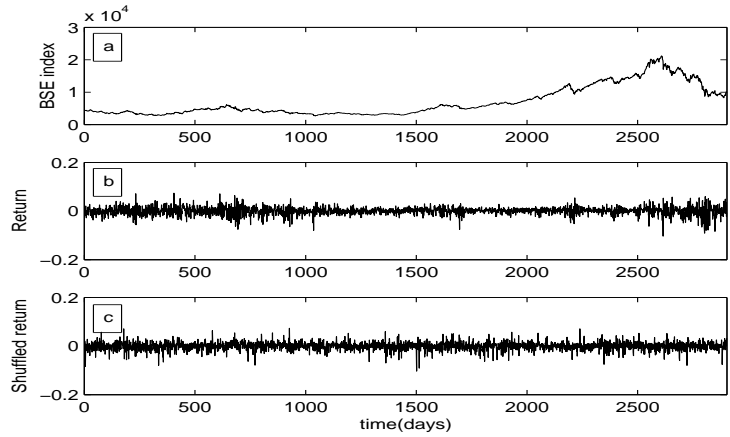}
\caption{\label{fig:BSE} (a) BSE high price index value in
daily trading over a period of 2903 days. (b)Logarithmic returns
estimated from Eq.\ref{eq:returns}. (c) Shuffled returns.}
\end{figure}

The BSE index \cite{yahoo} dates from July 01, 1997 to March 31,
2009. The data spans over 2903 points and is shown in
Fig.\ref{fig:BSE}(a). As is evident from the data, the first half
does not show much activity but the second half shows significant
variations. Fig.\ref{fig:BSE}(b) depicts the logarithmic returns
calculated from \eqref{eq:returns} and Fig.\ref{fig:BSE}(c) depicts
the shuffled returns, which reveals some differences with the returns.

\section{Continuous wavelet analysis through Morlet wavelet}
\label{sec:wt}
\begin{figure}
\subfigure[Scalogram]{
\includegraphics[width=5.5cm,height=5cm]{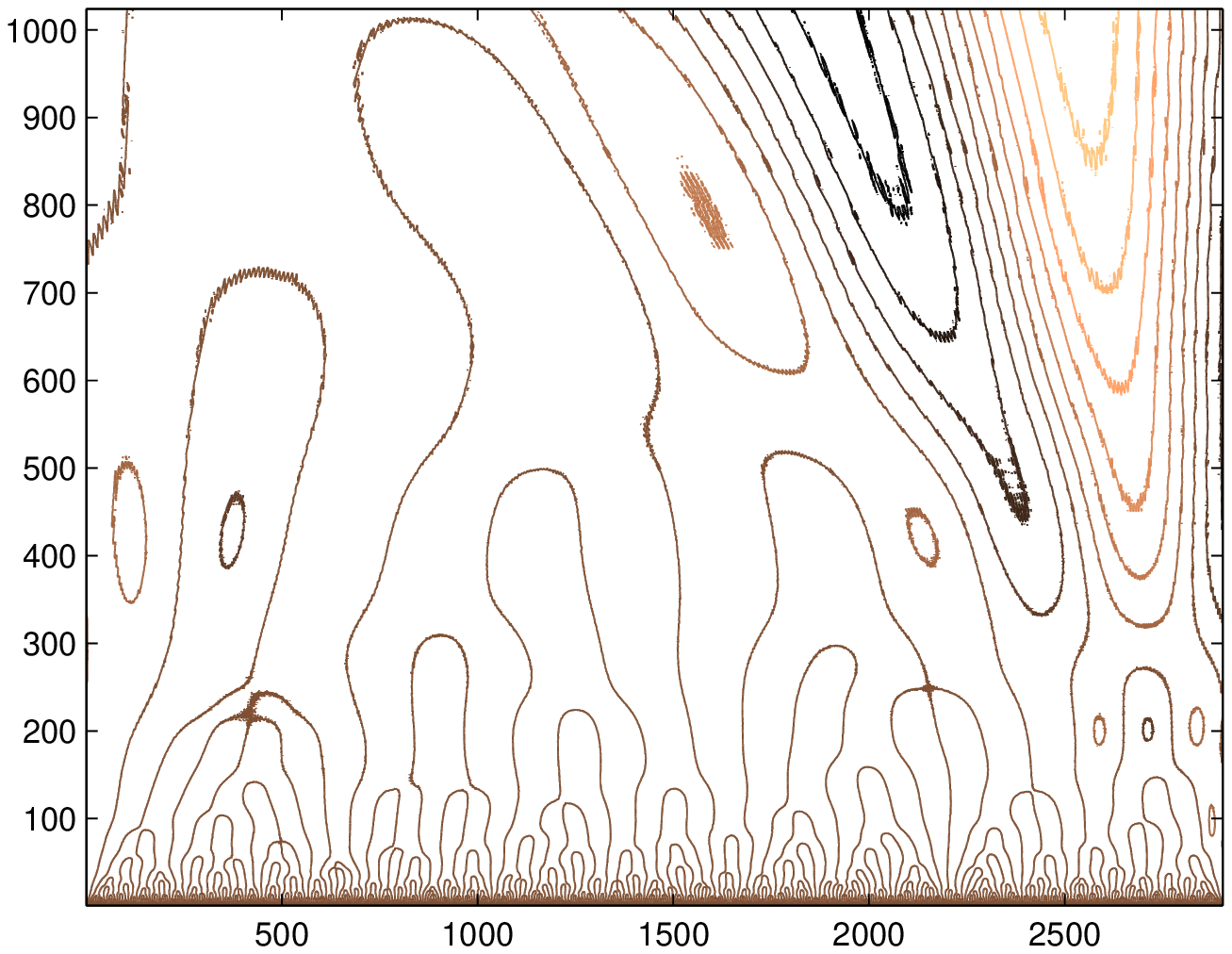}\label{fig:cont_cwt}
} \subfigure[Periodogram]{
\includegraphics[width=5.5cm,height=5cm]{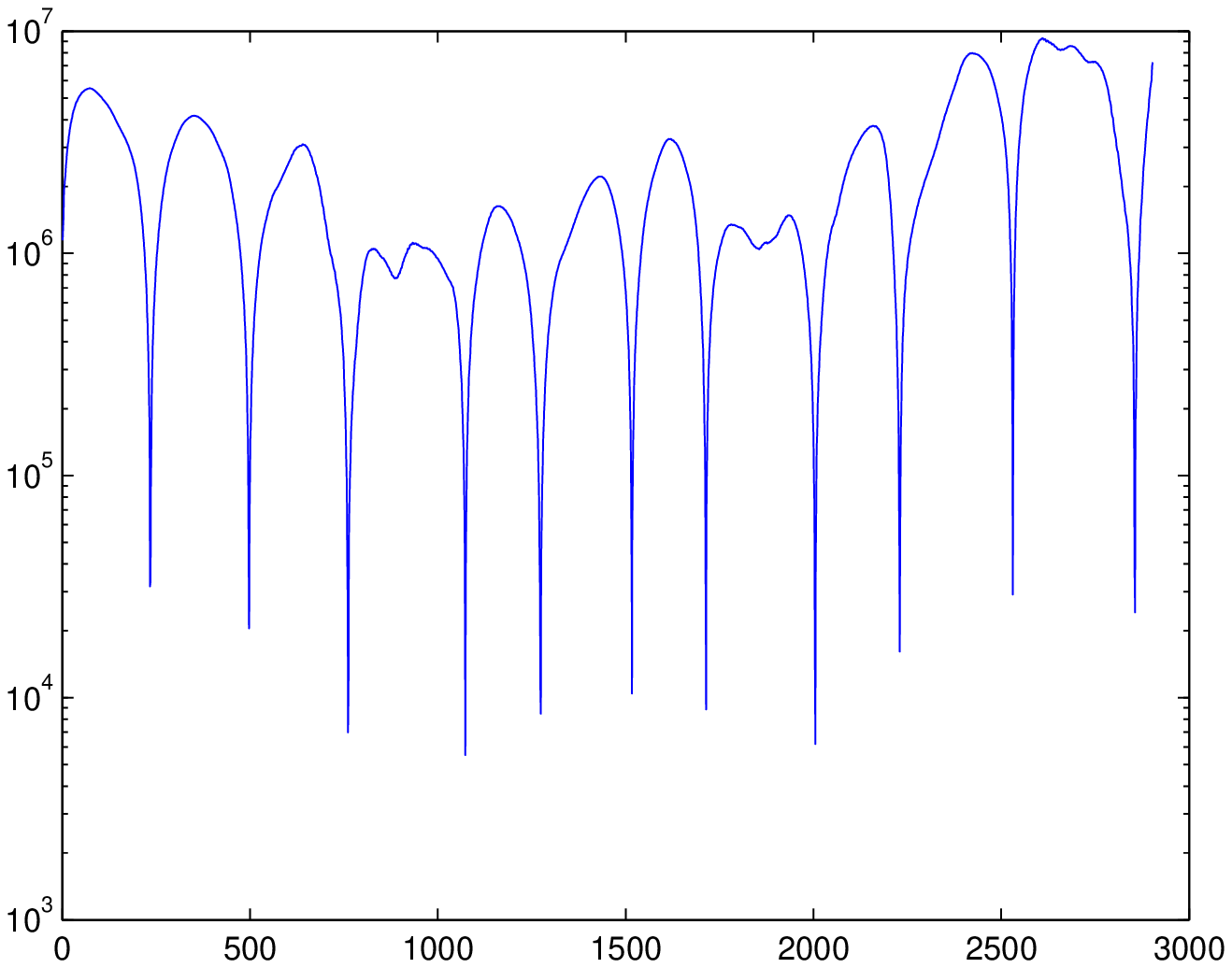}\label{fig:period_smlgy}
}
\label{continuous}
\caption{\subref{fig:cont_cwt} is the scalogram of the wavelet coefficients computed from scale 1 to 1024. The x-axis is the time, $n$ and the y-axis is the scale $s$. \subref{fig:period_smlgy} Periodogram  plotted on a semilog scale as $P_n$ vs $n$. One observes a period of approximately 250 trading days.}
\end{figure}

\begin{figure}
\includegraphics[width=11cm,height=8cm]{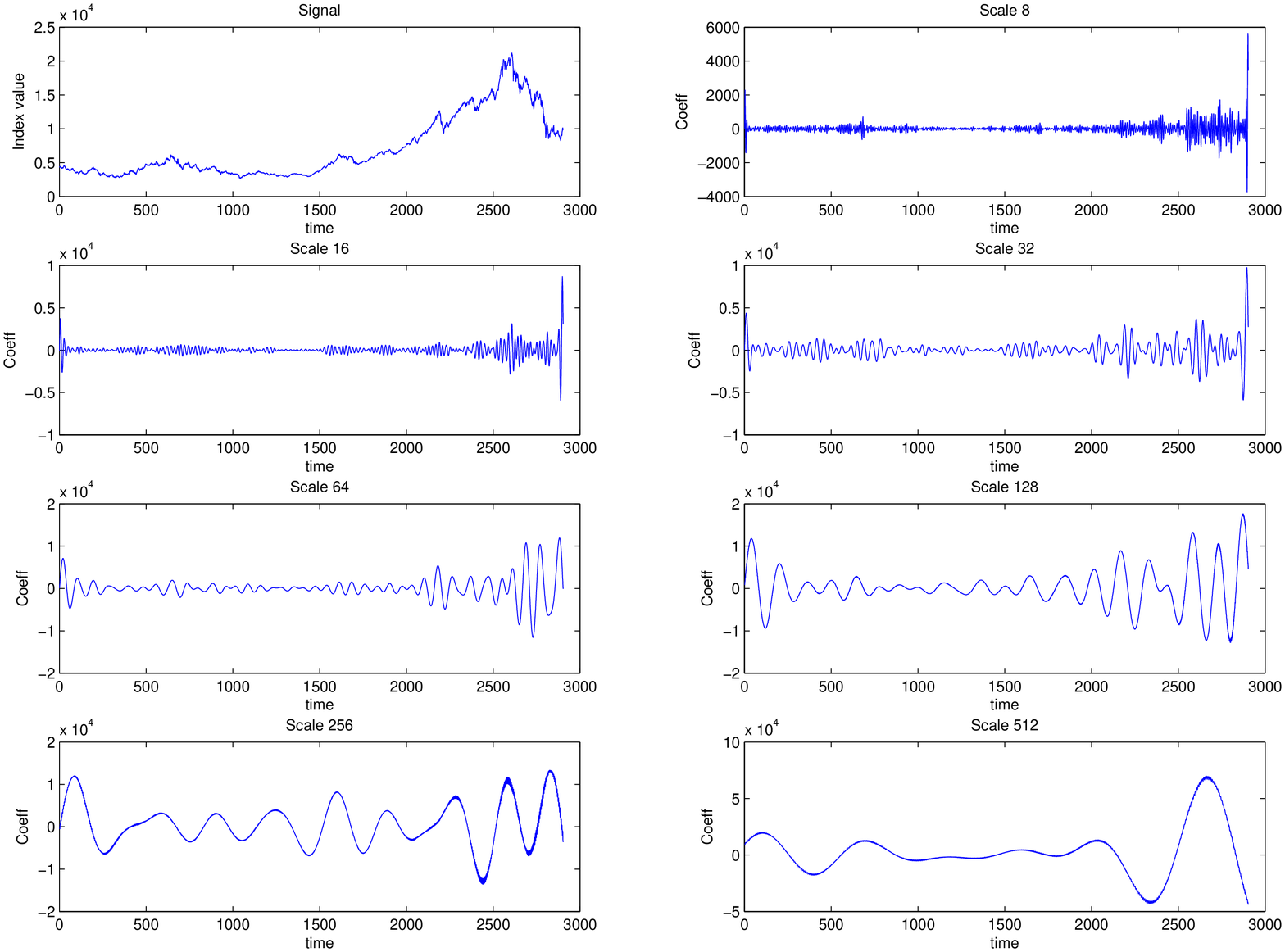}
\caption{\label{fig:period_scale}Wavelet coefficients at scales 8,
16, 32, 64, 128, 256 and 512.}
\end{figure}

Continuous Wavelet Transform (see \cite{farge} and \cite{torrence} for an excellent introduction
to the topic) has been used in recent times to analyze financial time series to study self-organized
criticality \cite{bartolozzi_1,bartolozzi_2}, correlations \cite{struzik,awc}, commodity prices \cite{connor}
to name a few. Recently, in \cite{dilip}, an effort towards the characterization of cyclic behavior in the financial
markets has been made through the multi resolution analysis of wavelet transforms. Here, the CWT of the BSE data has
been carried using the Morlet wavelet given by \cite{morlet},
\begin{equation}
\label{eq:morlet}
\psi_0 (n)=\pi^{-1/4}e^{\imath \omega_0 n}e^{-n^{2} /2}
\end{equation}
where $n$ is a localized time index, $\omega_0=6$ for zero mean and localization
in both time and frequency space (admissibility conditions for a wavelet) \cite{farge}.
The Morlet wavelet has a Fourier wavelength $\lambda$ \cite{torrence} given by
\begin{equation}
\label{fu_wavelength}
\lambda= \frac{4 \pi s}{\omega_0 +\sqrt{2+\omega_0^2}} \approx 1.03s
\end{equation}
which means that here, the scale and the Fourier
wavelength are approximately equal. The wavelet coefficients are
calculated \cite{torrence} by the convolution of a discrete sequence
$x_n$ with scaled and translated $\psi_0 (n)$,
\begin{equation}
\label{cwt} W_n(s)=\sum_{n'=0}^{N-1} x_{n'}\psi^{*}\left[
\frac{(n-n')}{s} \right]
\end{equation}
where $s$ is the scale. The wavelet coefficients for the BSE data
has been given in a scalogram in Fig.\ref{fig:cont_cwt} as a
function of scale and time. The periodicity of the coefficients over
the scales is calculated as
\begin{equation}
\label{eq:periodicity}
P_n=\sum_n W_n(s)
\end{equation}
and it is given in
Fig.\ref{fig:period_smlgy}.
To analyze the periodicity of the data at different frequencies,
\begin{equation}
\label{eq:sc_fr}
s \propto \nu^{-1},
\end{equation}
where $\nu$ is the frequency; we have shown the $W_n(s)$ at different scales in Fig.\ref{fig:period_scale}. 
One observes significant fluctuations at different scales in the second half of the data.It is evident that the fluctuations have a self similar character. We depict the fluctuations at smaller scale as also the dominant periodic variations at different scales and one does not see significant transient fluctuations in the variations.

\begin{figure}
\centering
\subfigure[DWT with Haar]{\includegraphics[width=5.5cm,height=6cm]{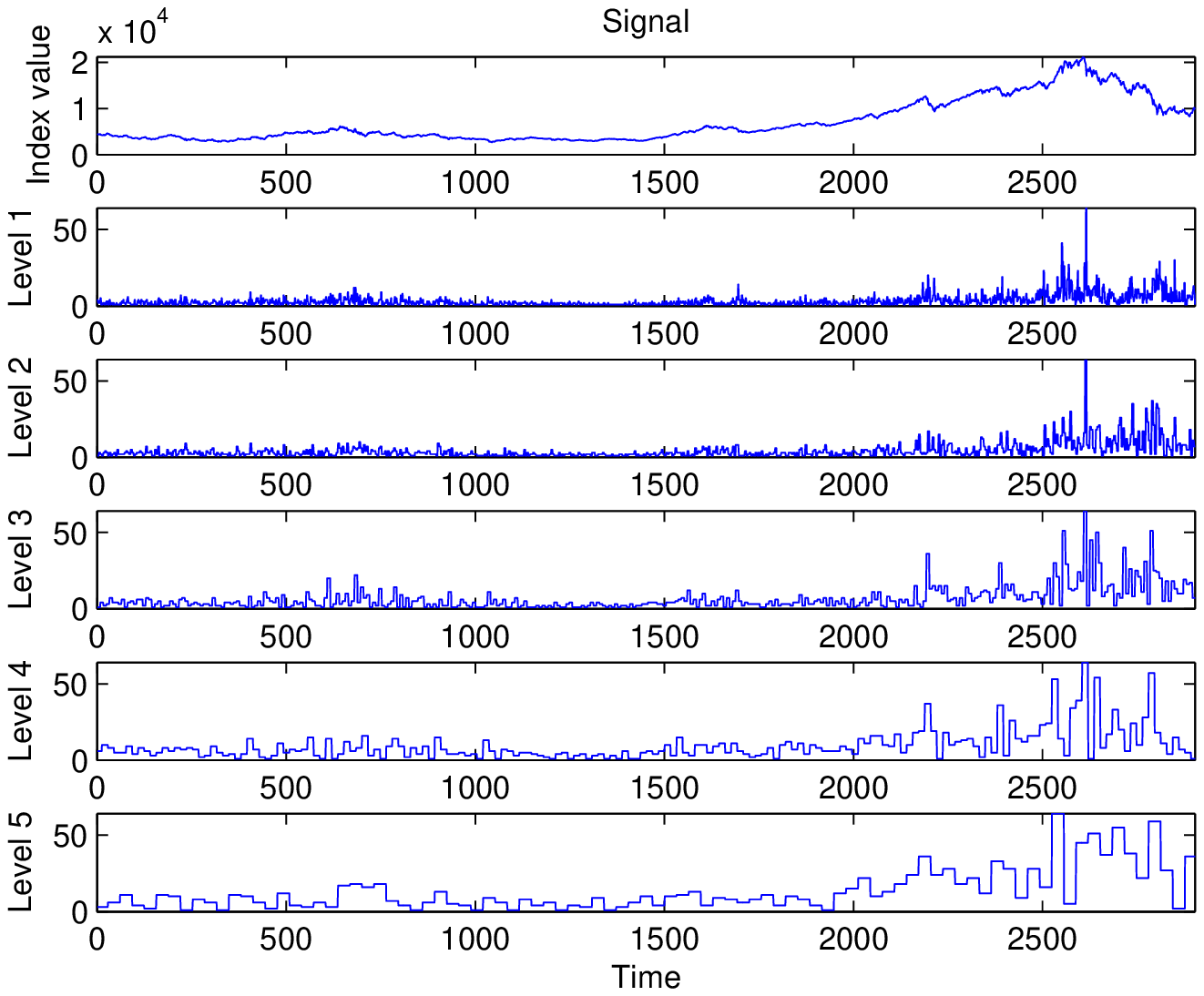}\label{fig:haar}}
\subfigure[DWT with Db-4]{\includegraphics[width=5.5cm,height=6cm]{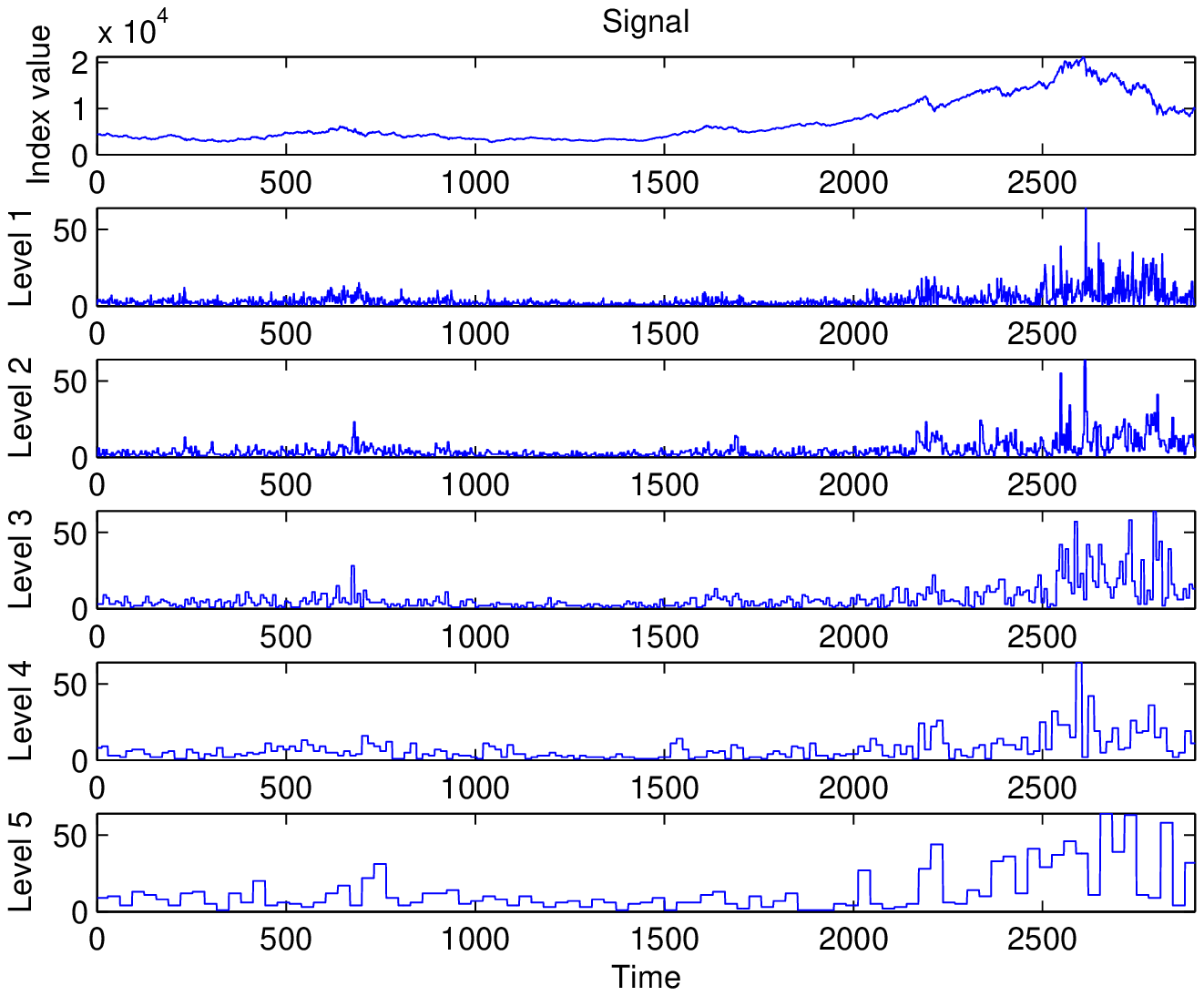}\label{fig:db4}}
\label{fig:dwt}
\caption{\subref{fig:haar} Discrete wavelet transform (DWT) of the data through Haar wavelet. \subref{fig:db4}DWT of the data through Daubechies-4 (Db-4) wavelet. Akin to the CWT case, the fluctuations show self similar behavior.}
\end{figure}

The extracted fluctuations at different levels through DWT are shown
in Fig.\ref{fig:haar} and Fig.\ref{fig:db4}. Having seen the
periodic behavior of the data, and having extracted the fluctuations
through DWT, in the next section, we discuss the wavelet based method
for analysis of fluctuations to identify their fractal behavior.

\section{Discrete wavelet based method for characterizing multifractal behavior}\label{sec:wbm}
We have observed earlier the
self-similar nature of the fluctuations in the wavelet domain. In the following we describe the procedure of the wavelet based method.

From the financial (BSE stock index) time series $x(t)$, the scaled
logarithmic returns $G(t)$ is defined as,

\begin{equation} \label{eq:returns}
G(t)\equiv \frac{1}{\sigma}[\log(x(t+1))- \log(x(t))],\qquad t=1,2...(N-1);
\end{equation}

here $\sigma$ is the standard deviation of $x(t)$. The profile of
the time series is obtained from the cumulative,
\begin{equation}
Y(i) = \sum_{t=1}^i [G(t)],\qquad i=1,....,N-1.
\end{equation}

Next, apply the wavelet transform on the time series profile $Y(i)$
to extract the fluctuations from the trend. The trend is extracted
by discarding the high-pass coefficients and reconstructing only
with low-pass coefficients using inverse wavelet transform. The
fluctuations are then extracted at each level by subtracting the
trend from the original time series. This procedure is followed to
extract fluctuations at different levels. Here the wavelet window
size at each level of decomposition is considered as the scale $s$.
We have made use of Daubechies (Db) wavelets for the extraction of
desired polynomial trend. Although the Daubechies wavelets extract
the fluctuations effectively, its asymmetric nature and wrap around
problem affects the precision of the values. We apply wavelet
transform on the reverse profile, to extract a new set of
fluctuations. These fluctuations are then reversed and averaged over
the earlier obtained fluctuations.

Now the extracted fluctuations using wavelet transform are
subdivided into non-overlapping segments $M_s = int(N/s)$ where $N$
is the length of the fluctuations and $s$ is the scale. The $q^{th}$
order fluctuation function $F_q(s)$ is then obtained by squaring and
averaging the fluctuations over all segments:
\begin{equation}
F_q(s) \equiv  \left( \frac {1}{2 M_s} \sum_{b=1}^{2 M_s} [
F^2(b,s)]^{q/2}\right)^{1/q}.
\end{equation}

Here '$q$' is the order of moment. The above procedure is repeated
for different scale sizes for different values of $q$ (except
$q=0$). The power law scaling behavior is obtained from the
fluctuation function,

\begin{equation}
F_q(s) \sim s^{h(q)},
\end{equation}
in a logarithmic scale for each value of $q$. If the order $q = 0$,
direct evaluation leads to the divergence of the scaling
exponent. In that case, logarithmic averaging has to be employed to
find the fluctuation function:
\begin{equation}
F_q(s) \equiv exp \left( \frac {1}{2 M_s} \sum_{b=1}^{2 M_s} ln [
F^2(b,s)]^{q/2} \right)^{1/q}.
\end{equation}

For the monofractal time series, $h(q)$ values are independent of q
and for the multi-fractal time series $h(q)$ values are dependent on $q$. $h(q=2)=H$, the Hurst scaling exponent is a measure of
fractal nature such that varies $0 < H < 1$. Here $H < 0.5$ and $H > 0.5$ reveal the anti-persistent and persistent nature of the time
series, whereas $H=0.5$ is for random time series.

\section{Data analysis and observations}\label{sec:obs}
\begin{figure}
\centering 
\subfigure[]{\includegraphics[scale=0.4]{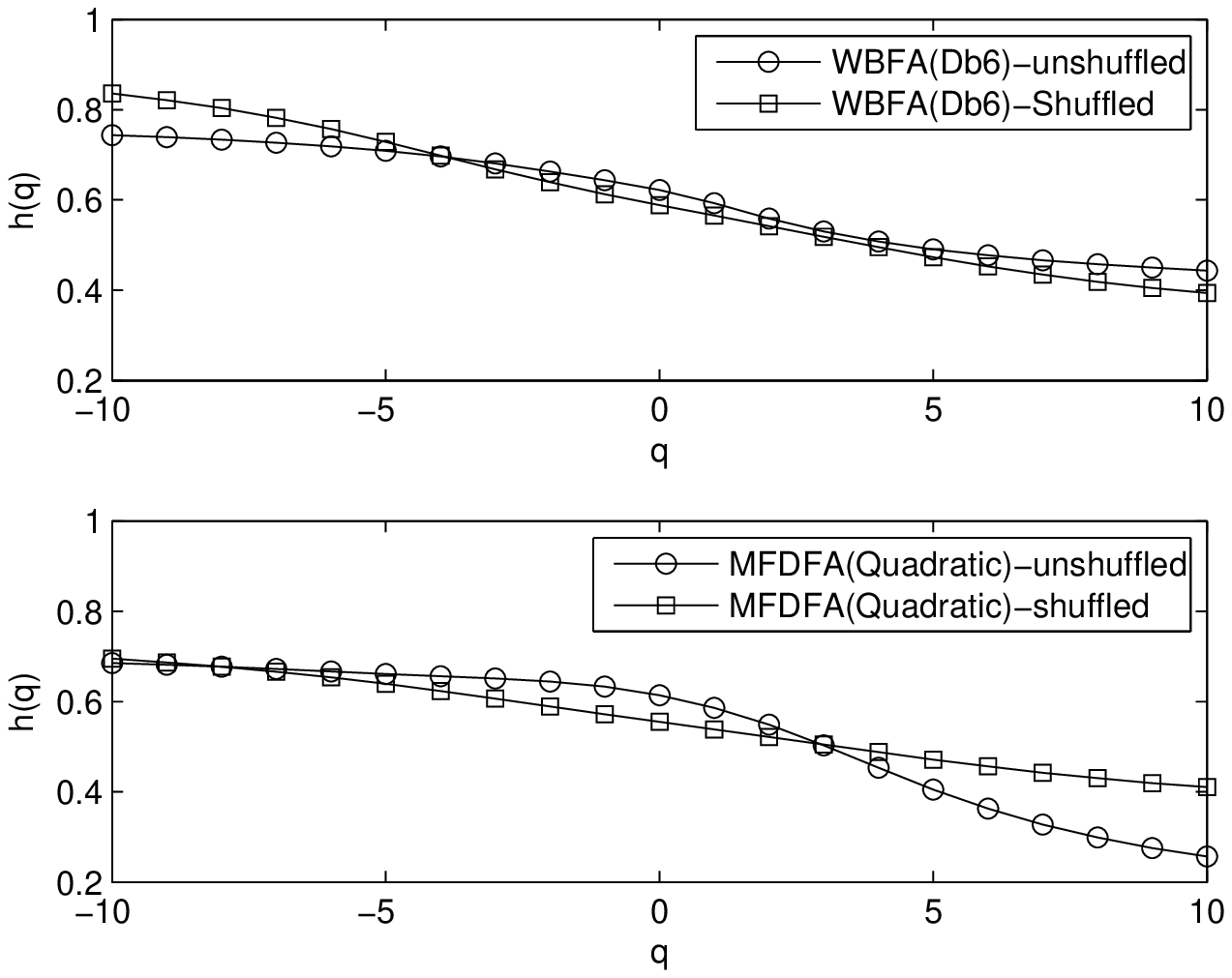}\label{fig:BSE_hq}}
\subfigure[]{\includegraphics[scale=0.35]{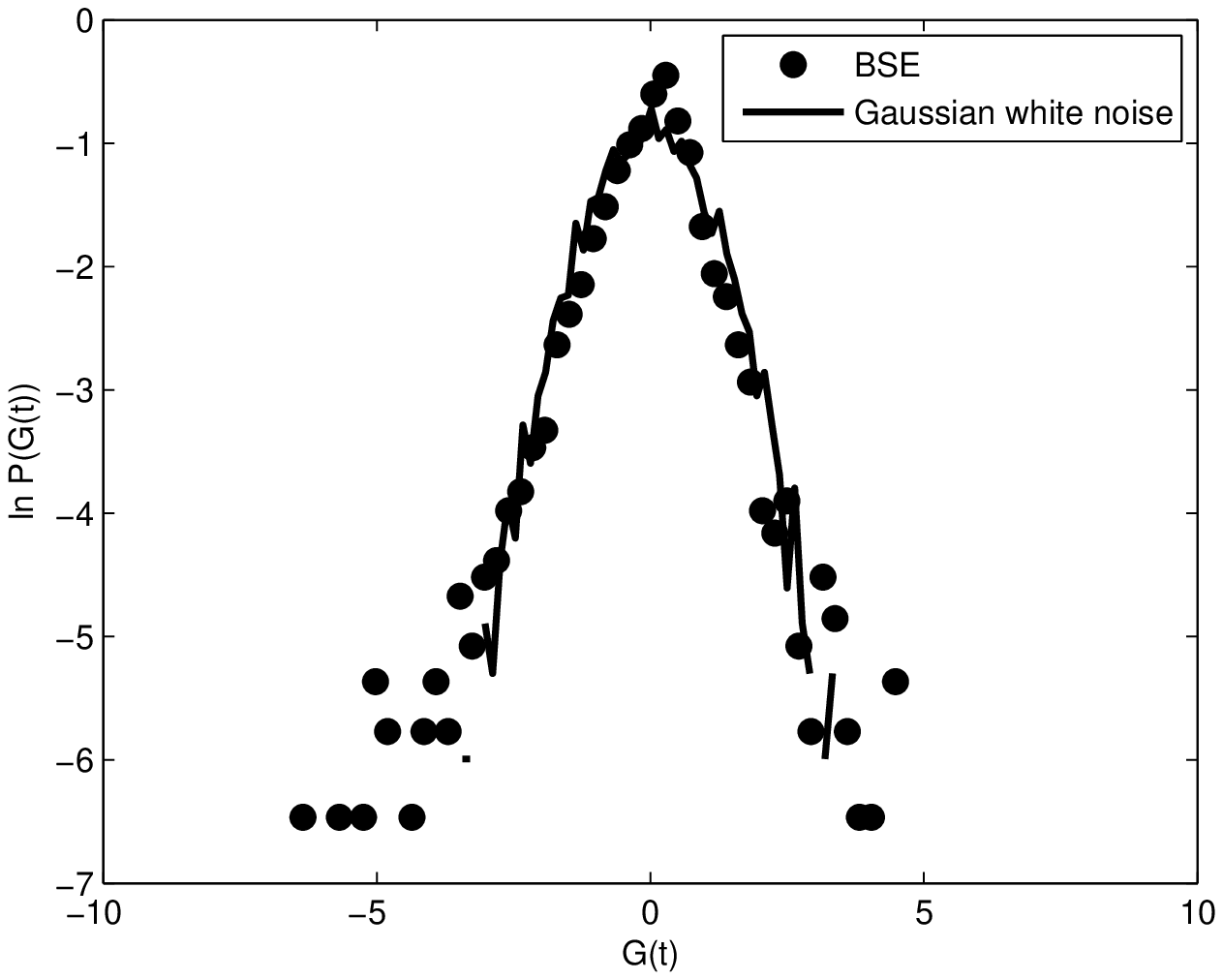}\label{fig:BSE_Log}}
\label{fig:hq_log}
\caption{\subref{fig:BSE_hq} $h(q)$ values of BSE Sensex price index using (Top panel) WBFA (Db-6) and (Bottom panel) MF-DFA (Quadratic) analysis. \subref{fig:BSE_Log} Log Normal Distribution of BSE sensex index return and Gaussian white noise.}
\end{figure}

The wavelet based fluctuation analysis (WBFA) which is used here was carried on the time series profile obtained from the returns and
shuffled returns. The analyzed time series using discrete wavelet
based method with Db-6 wavelet reveals the presence of multifractal
nature with long-range correlation behavior that is shown in Fig.\ref{fig:BSE_hq} (top panel). For the sake of comparison, MF-DFA method with
quadratic polynomial fit is also used which complements the
wavelet based method (see Fig.\ref{fig:BSE_hq} (bottom panel)). The Hurst scaling
exponent reveals that the time series possesses persistent behavior, which is shown in Table \ref{tab:BSE_table}. The semi-log plot of distribution of
logarithmic returns of BSE index and the Gaussian white noise is
shown in Fig. \ref{fig:BSE_Log} for the BSE index. The fat tails for large
fluctuations and sharper behavior near the origin for small
fluctuations are clearly seen.

\begin{figure}
\centering
\subfigure{\includegraphics[width=3.5cm,height=4cm]{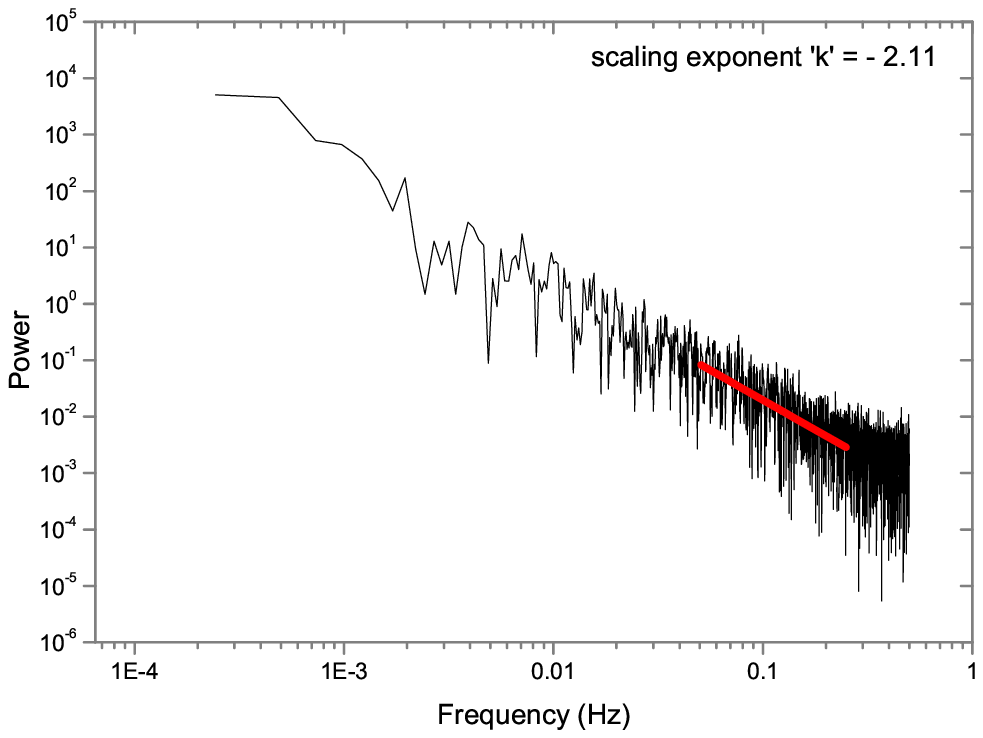}\label{fig:BSE_FFT}}
\subfigure{\includegraphics[width=3.5cm,height=4cm]{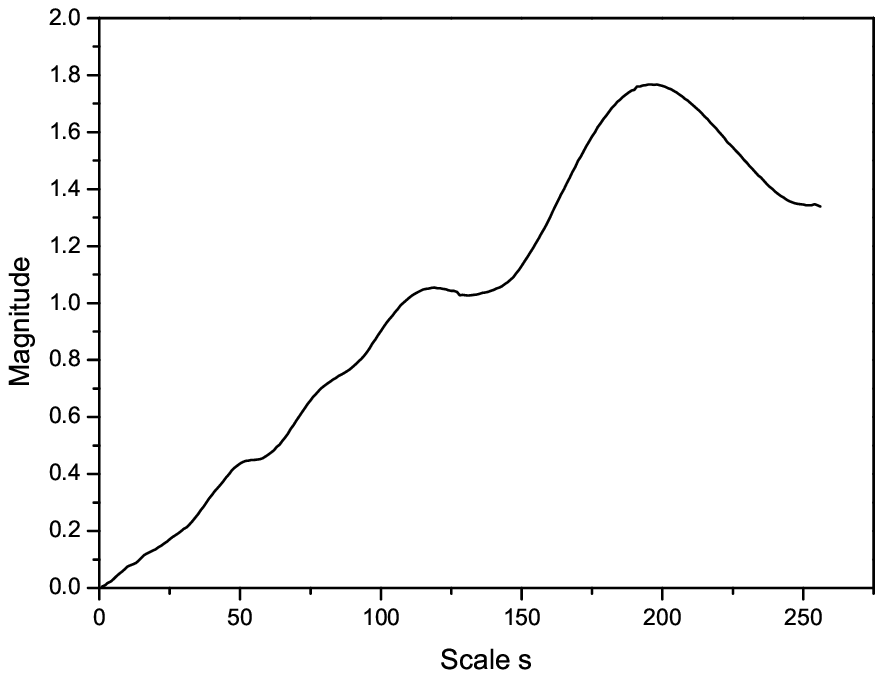}\label{fig:CWT_mag_scale}}
\subfigure{\includegraphics[width=3.5cm,height=4cm]{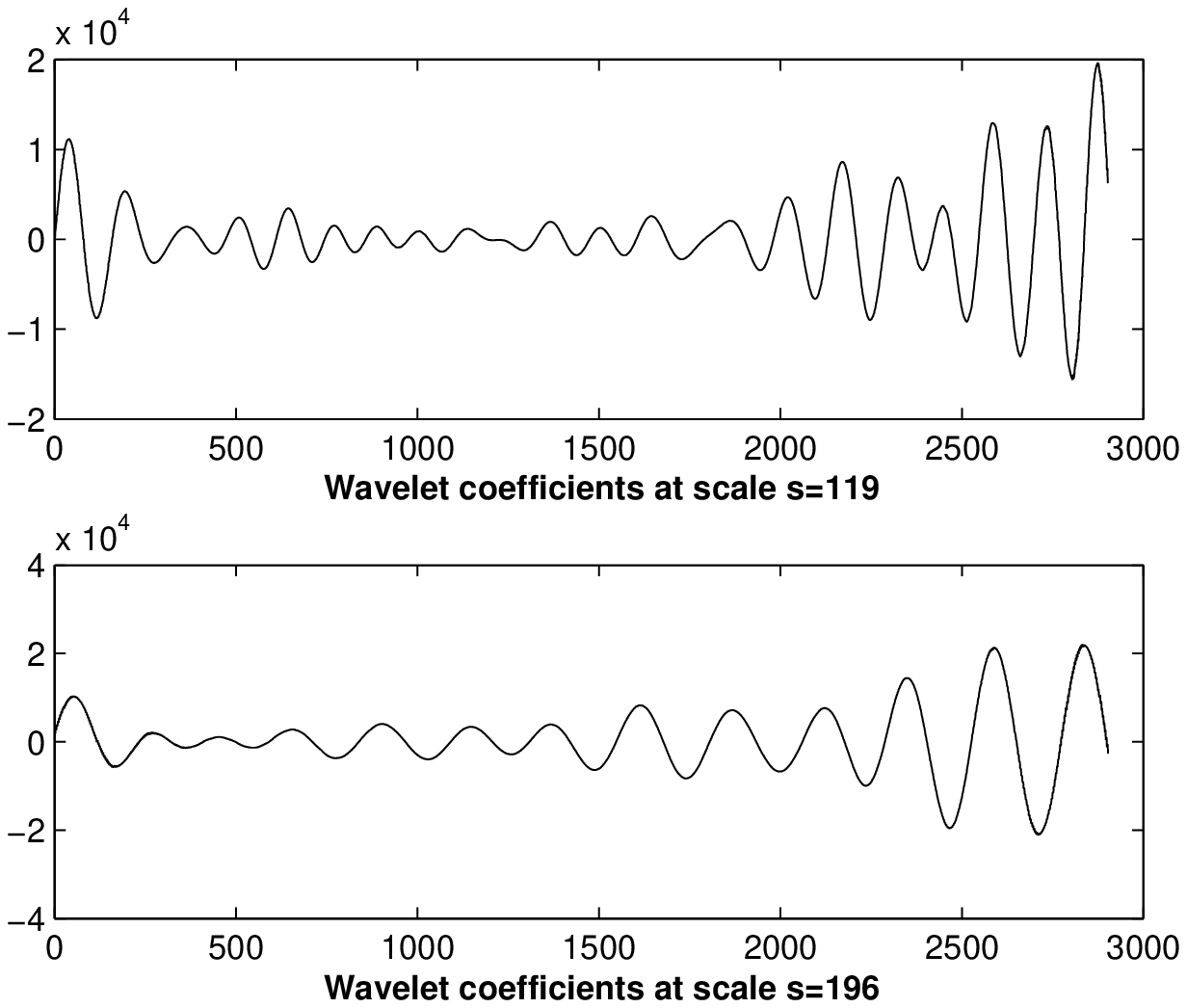}\label{fig:CWT_scale}}
\caption{\subref{fig:BSE_FFT} Fourier power spectral analysis on BSE price index. \subref{fig:CWT_mag_scale} Continuous wavelet analysis show two dominant periodic modulations at scale 119 and 194. \subref{fig:CWT_scale} The CWT coefficients at the above scales (119 and 194) as a function of time.}
\end{figure}

\begin{table}
\begin{tabular}{|c|c|c|c|c|}
\hline
X & $h(q)_{WBFA}$ & $h(q)_{WBFA_{s}}$ & $h(q)_{MFDFA}$ &$h(q)_{MFDFA_{s}}$ \\
\hline
Hurst Scaling Exponent & 0.5486 &0.5218 &0.5590 &0.5420\\
\hline \hline
\end{tabular}
\caption{\label{tab:BSE_table} $h(q)$ versus $q$ values for WBFA (Db-6) and MFDFA (Quadratic) analysis.}
\end{table}
We have also analyzed the scaling behavior through Fourier power
spectral analysis,
\begin{equation}
P(s) = \left | \int Y(t)e^{-2\pi \imath s t} dt\right |^2.
\end{equation}
Here $Y(t)$ is the accumulated fluctuations after subtracting the
mean $\langle Y \rangle$. It is well known that,  $P(s) \sim s^{-
\alpha}$.  For the BSE price index time series, the scaling exponent
$\alpha = 2.11$ which reveals long range correlated behavior as
shown in Fig. \ref{fig:BSE_FFT}. The obtained scaling exponent $\alpha$ can be
compared with Hurst exponent by the relation $\alpha = 2 H + 1$. The
wavelet based method and FFT are comparable.

\section{Conclusion}\label{sec:disc}

We have analyzed BSE high price index values in daily trading. A
detailed study reveals multifractal behavior and
non statistical distribution of the returns. The distribution function of the
returns also show fat-tail behavior. The analysis through the wavelet based
method, MFDFA method and also Fourier power spectrum analysis reveal a persistent
nature, as well as multifractal behavior of the BSE price index values. The multifractal nature of the time series may arise due to herding behavior, and other intrinsic non-linear character of the market and other control mechanism. We intend to study the fluctuations in different price indices in different countries for this time period. This may reveal the physical origin of the other time periods as also the multi fractal character.

\section{Acknowlegement}
PM, one of the authors would like to thank the Department of Science and Technology for their financial support (DST-CMS GoI
Project No. SR/S4/MS:516/07 Dated 21.04.2008).


\begin{thebibliography}{}
\bibitem{plerou}Plerou V \textit{et. al} (2000), \textit{Physica A} {\bf 279}:443.
\bibitem{bachelier}Bachelier L (1900) \textit{Ann. Sci. \'Ecole Norm. Sup.} {\bf 3}:21.
\bibitem{pareto}Pareto V (1897) \textit{Cours d'\'Economie Politique Lausanne}, Paris.
\bibitem{levy}L\'evy P (1937) \textit{Th\'eorie de l'Addition des Variables Al\'e atoires, Gauthier-Villars}, Paris.
\bibitem{mandelbrot_63}Mandelbrot B B (1963) \textit{J. Bus.} {\bf36}:394419.
\bibitem{mantegna_2000} Mantegna, R N, Stanley H E (2000), \textit{Introduction to Econophysics: Correlations and Complexity in Finance}, Cambridge University Press, Cambridge.
\bibitem{bouchad}Bouchaud J P, Potters M (2000) \textit{Theory of Financial Risk}, Cambridge University Press, Cambridge.
\bibitem{farmer}Farmer J D (1999) \textit{Comput. Sci. Eng.} {\bf 1}:26.
\bibitem{kondor}Kondor I, K\'ertesz (eds.) (2000) \textit{Econophysics: An Emerging Science}, Kluwer, Dordrecht.
\bibitem{mantgna_99}Mantegna R N, (ed.) (1999) \textit{Proceedings of the International Workshop on Econophysics and Statistical Finance, Physica A (special issue)} {\bf 269}:1.
\bibitem{bouchad_2000}Bouchaud J P, Alstr\"om P, Lauritsen K B (eds.), (2000) \textit{Application of Physics in Financial Analysis, Int. J. Theor. Appl. Finance (special issue)} {\bf 3}.
\bibitem{takayasu}Takayasu H (ed.) (2002), \textit{The Application of Econophysics: Proceedings of the Second Nikkei Econophysics Symposium},Springer.
\bibitem{mandel}Mandelbrot B B (1999),{\it The Fractal Geometry of Nature} Freeman, San Francisco.
\bibitem{daub}Daubechies I (1992) {\it Ten lectures on wavelets} SIAM, Philadelphia.
\bibitem{mall}Mallat S (1999) {\it A Wavelet Tour of Signal Processing} Academic Press.
\bibitem{ram}Burrus C S, Gopinath R A, and Guo H (1998) {\it Introduction to Wavelets and Wavelt Transforms} Prentice Hall, New Jersey.
\bibitem{mani1}Manimaran P, Panigrahi P K, and Parikh J C (2005), {\it Phys. Rev. E} {\bf 72}:046120.
\bibitem{mani3}Manimaran P, Lakshmi P A, Panigrahi P K (2006), \textit{J. Phys. A} {\bf 39}:L599.
\bibitem{drozdz}O\'swiecimka P, Kwapi\'en, Drozdz (2006) \textit{Phys. Rev. A.} {\bf 74}:016103.
\bibitem{mani2}Manimaran P, Panigrahi P K, and Parikh J C (2008) {\it Physica A} {\bf 387}:5810.
\bibitem{mani4}Manimaran P, Panigrahi P K, and Parikh J C (2009) {\it Physica A} {\bf 388}:2306.
\bibitem{khu}Hu K, Ivanov P Ch, Chen Z, Carpena P, and Stanley H E (2001) {\it Phys.Rev. E} {\bf 64}:11114.
\bibitem{gopi}Gopikrishnan \textit{et. al} (1999) {\it Phys. Rev. E} {\bf 60}:5305.
\bibitem{ple}Plerou V \textit{et. al} (1999) {\it Phys. Rev. E} {\bf 60}:6519.
\bibitem{chen}Chen Z, Ivanov P Ch., Hu K, and Stanley H E (2002) {\it Phys. Rev. E} {\bf 65}:041107.
\bibitem{matia}Matia K, Ashkenazy Y, and H. E. Stanley, (2003) {\it Europhys. Lett.} {\bf 61}:422.
\bibitem{hwa}Hwa R C \textit{et. al}(2005) \textit{Phys. Rev. E.} {\bf 72}:066308.
\bibitem{ohashi}Ohashi K, Amaral L A N, Natelson B H, Yamamoto Y (2003) \textit{Phys. Rev. E.} {\bf 68}:065204.
\bibitem{xu}Xu L \textit{et. al}(2005) \textit{Phys. Rev. E.} {\bf 71}:051101.
\bibitem{brodu}Brodu N, eprint:nlin.CD/0511041.
\bibitem{gu}Gu G F, Zhou W X (2006) \textit{Phys. Rev. E} {\bf 74}:061104.
\bibitem{mohaved}Mohaved M S, Hermanis E (2008) \textit{Physica A} {\bf 387}:915.
\bibitem{yahoo}obtained from http://in.finance.yahoo.com
\bibitem{farge}Farge M, (1992) \textit{Annu. Rev. Fluid Mech.} {\bf 24}:395.
\bibitem{struzik}Stuzik Z R, (2001) \textit{Physica A} {\bf 296}:307.
\bibitem{bartolozzi_1}Bartolozzi M \textit{et. al} (2005) \textit{Physica A} {\bf 350}:451.
\bibitem{bartolozzi_2} BartolozziM (2007) \textit{Eur. Phys. J. B} {\bf 57}:337.
\bibitem{awc}Simonsen I, Hansen A, and Nes O -M, \textit{Phys. Rev. E} {\bf 58} (1998) 2779.
\bibitem{connor}Connor J and Rossiter R (2005), \textit{Studies in Nonlinear Dynamics and Econometrics} {\bf 9}:1.
\bibitem{dilip}Ahalpara D P \textit{et. al} (2008) \textit{Pramana -J. Phys.} {\bf 71}:459.
\bibitem{torrence}Torrence C and Compo G P (1998) \textit{Bull. Amer. Meteorol. Soc.} {\bf 79}:61.
\bibitem{morlet}Goupillaud P, Grossman A, and Morlet J (1984) \textit{Geoexploration} {\bf 23}:85-102.
\bibitem{hur}Hurst H E (1951) {\it Trans. Am. Soc. Civ. Eng.} {\bf 116}:770.
\bibitem{feder}Feder J(1988) {\it Fractals} Plenum Press, New York,
\bibitem{arn1}Arneodo A \textit{et. al} (1988) {\it Phys. Rev. Lett.} {\bf 61}:2284; Muzy J F \textit{et. al} (1993) {\it Phys. Rev. E} {\bf47}:875.
\bibitem{peng}Peng C K, \textit{et. al} (1994) {\it Phys. Rev. E} {\bf 49}:1685.
\bibitem{net}Kantelhardt J W, \textit{et. al} (2003) {\it Physica A} {\bf 330}:240.
\end{thebibliography}
\end{document}